\documentclass[12pt, letter]{article}

\usepackage[margin=1in]{geometry}

\usepackage[utf8]{inputenc}
\usepackage{microtype}
\usepackage{setspace}

\usepackage{amsmath}
\usepackage{amsfonts}
\usepackage{amssymb}

\usepackage{graphicx}
\usepackage{booktabs}
\usepackage{caption}
\usepackage{floatrow}
\usepackage{siunitx}
\usepackage{makecell}
\usepackage{threeparttable}

\usepackage{algorithm}
\usepackage{algpseudocode}

\usepackage[round]{natbib}

\newcommand{\rev}[1]{#1}

\usepackage[colorlinks=true,
    linkcolor=black,
    urlcolor=black,
    citecolor=black]{hyperref}

\usepackage{array}

\title{\textbf{An Auditable AI Agent Loop \\ for Empirical Economics: \\A Case Study in Forecast Combination} \\$~$}

\author{
  Minchul Shin \\ \textit{Federal Reserve Bank of Philadelphia}\thanks{The views expressed herein are those of the author and do not necessarily reflect the views of the Federal Reserve Bank of Philadelphia or the Federal Reserve System. Email: visiblehand@gmail.com.}
  \\ $~$
  \\ $~$
}
\date{First version: March 17, 2026 \\ This version: \today}

\begin{document}

\newgeometry{top=0.6in,bottom=0.6in,left=1in,right=1in}

\maketitle

\doublespacing

\begin{abstract}
AI coding agents\rev{, general purpose assistants that write and execute code,} make empirical specification search fast and cheap, but they also widen hidden researcher degrees of freedom. This paper adapts an open-source agent-loop architecture to an empirical economics workflow and adds a post-search holdout evaluation. In a forecast-combination illustration, independent agent searches find methods that improve on benchmarks from the original study. Logged search and holdout evaluation together make adaptive specification search more transparent and help distinguish robust improvements from sample-specific discoveries.
\end{abstract}

\vspace{1cm}

\newpage
\restoregeometry
\doublespacing


\section{Introduction}
\label{sec:intro}

Empirical researchers have always explored alternative specifications; AI coding agents\rev{, general purpose assistants such as Claude Code, Codex, or OpenCode that write and execute code,} make that exploration cheap and fast. \rev{The change is not only speed. Conventional specification search usually ranges over a menu fixed in advance, such as regressors or functional forms in $y = x'\beta + u$. Agentic search instead ranges over code. The researcher fixes the inputs, the object to be produced, and the score: for example data $(y,x)$, predictions $\hat y$, and out-of-sample MSE. The agent can then write code that maps the inputs into predictions, such as $\hat y = g(x;\hat\beta)$. The searched object is therefore the code itself, not only a pre-listed formula or tuning parameter.}

\rev{Researcher degrees of freedom widen accordingly, and} the boundary between coding assistance and undisclosed specification search becomes thin. For a field long concerned with such degrees of freedom \citep{Leamer1983,White2000,GelmanLoken2013,Miguel2021}, the central question is not whether AI agents can improve an empirical score, but whether their search can be made transparent and auditable.

Building on \citeauthor{Karpathy2026}'s (\citeyear{Karpathy2026}) open-source \texttt{autoresearch}, this paper adapts a simple agentic coding loop into a research protocol for empirical economics. The protocol has four ingredients: a written instruction contract, an immutable evaluator, a single editable script, and a complete experiment log. This paper also adds an explicit post-search holdout evaluation. Together, these elements make agentic search auditable: they record what was tried, limit where and how much the agent can search, and test the result beyond the search sample.

The empirical illustration applies this agentic-search protocol to the forecast-combination problem studied by \citet{DieboldShin2019}. The agent searches for improvements to the method proposed there. On the original rolling evaluation, three independent search runs beat existing methods. A post-search holdout for 2017Q1--2025Q4 then evaluates the discovered rules: two continue to outperform standard benchmarks, while one does not. The logs show how each search moved from the initial task toward broader method discovery; the holdout shows which discoveries survive outside the search sample.

Recent AI-for-science work ranges from code rewriting against automated evaluators \citep{NovikovEtAl2025,ChenEtAl2025} to broader research automation \citep{LuEtAl2024,GottweisEtAl2025}; in economics, \citet{Korinek2025} surveys AI agents, and \citet{DawidEtAl2025} develop agentic workflows. We study evaluator-locked local search in a fixed workflow.

\section{An auditable agent loop, in practice}
\label{sec:protocol}

\rev{The protocol is easiest to see as a small modification of ordinary empirical work. Hold fixed the empirical design: the data, sample, information set, and scoring rule. Put the specification choice in one script. A researcher could now edit that script, rerun the fixed pipeline, compare the new score with the old one, and keep the change if it helps. The agent loop automates exactly this edit--evaluate--compare routine. Its output is not only the best script found, but the full trail of scripts tried and scores obtained. Algorithm~\ref{alg:loop} writes this routine in notation; the next subsection maps that notation to four working-directory files.}

\begin{algorithm}[t!]
\caption{Auditable agent-loop protocol}
\label{alg:loop}
\begin{algorithmic}[1]
\Require Instruction contract $C$, evaluator $S(\cdot\,; D^{S})$, search sample $D^{S}$, initial script $\tau_0$, budget $K$
\State Run evaluator on $\tau_0$; record score; set $\tau^{\mathrm{best}} \leftarrow \tau_0$
\For{$k = 1, \ldots, K$}
  \State Agent proposes candidate script $\tau_k$, informed by log $L_{k-1}$
  \State Run evaluator on $\tau_k$; record score (or crash)
  \State If score improves, set $\tau^{\mathrm{best}} \leftarrow \tau_k$; otherwise revert to $\tau^{\mathrm{best}}$
  \State Append $(k,\, \tau_k,\, s_k,\, \text{outcome})$ to log $L_{k}$
\EndFor
\end{algorithmic}
\end{algorithm}

\subsection{From empirical workflow to four files}
\label{sec:fourfiles}

\rev{The notation in Algorithm~\ref{alg:loop} is implemented through four working-directory objects. The instruction contract $C$ is the researcher's written design document and, operationally, the prompt read by the coding agent. It states the objective, context, rules, and experiment budget. The editable script $\tau$ contains the specification being searched. The immutable evaluator $S(\tau;D^{S})$ holds fixed the empirical design, including the search sample $D^{S}$, and returns the score. The log $L$ records every attempt. Table~\ref{tab:files} maps these objects to files, first through a familiar regression-specification search and then through the forecast-combination application studied below.}

\begin{table}[t!]
\centering
\begin{threeparttable}
\caption{The four files of the protocol}
\label{tab:files}
\footnotesize
\rev{\begin{tabular}{@{}p{3.5cm}p{5.6cm}p{5.6cm}@{}}
\toprule
File & Regression example & This paper's application \\
\midrule
Instruction contract $C$ (\texttt{program.md}) & Define the search: improve validation fit in $y_i = x_i'\beta + g(z_i) + u_i$; edit only the code for $g$; hold the sample and scoring rule fixed; stop after $K$ attempts. & Find a tuning-parameter selection rule for peLASSO; edit only \texttt{train.R}; soft budget of roughly 200 experiments. Verbatim copy archived in Appendix~\ref{sec:app_program}. \\[10pt]

\midrule

Editable script $\tau$ (\texttt{train.R}) & The specification being searched: controls, transformations, interactions, shrinkage rules, or other code producing $\hat g(z_i)$. Each version is one candidate specification. & A function that maps the forecaster panel available at each forecast origin into one combined forecast using peLASSO with the candidate tuning-parameter selection rule. \\[10pt]

\midrule 

Immutable evaluator $S$ (\texttt{prepare.R}) & Holds fixed the empirical design. It runs each candidate script on the same data split and returns one score, such as validation MSE. & Calculates out-of-sample RMSE for the current \texttt{train.R} on the fixed search sample, passing only past data to \texttt{train.R}. \\[10pt]

\midrule

Experiment log $L$ (\texttt{results.tsv}) & Records every specification tried, its score, and whether it was kept, discarded, or crashed. & A tab-separated values file with about 200 rows, as detailed in Section~\ref{sec:empirical}. \\
\bottomrule
\end{tabular}}
\begin{tablenotes}\small
\item \rev{\textit{Notes:} The file extensions are incidental. \texttt{.md} is a Markdown text file; any readable text format, such as \texttt{.txt}, could serve as the instruction contract. \texttt{.R} is an R program; it could instead be \texttt{.py}, \texttt{.m}, \texttt{.do}, or any runnable script in the researcher's environment. \texttt{.tsv} is a tab-separated table; any editable table format, such as \texttt{.csv}, could serve as the log.}
\end{tablenotes}
\end{threeparttable}
\end{table}

\rev{This organization makes the adaptive search auditable by answering three questions. First, what was allowed to change? The file structure makes key restrictions machine checkable: the agent may edit \texttt{train.R}, but not the evaluator, and the evaluator controls which data are passed to the script. Any attempt to change another part of the codebase can be detected mechanically and treated as a failed experiment. Second, what was tried? Every candidate receives a mechanical status, \textit{keep}, \textit{discard}, or \textit{crash}, and remains in the log even if it is later reverted.\footnote{In our application, each candidate is also preserved as a git commit, so logged rows can be recomputed after the fact rather than trusted.} Third, how did the search arrive at the final script? Because the log is ordered and later proposals are informed by earlier successes and failures, the researcher can inspect the path from the baseline script to the selected one, including where the search improved, stalled, or drifted.}

\rev{The separation between immutable search design and editable implementation parallels a pre-analysis plan, except that the audit trail is automatic and complete. The architecture applies whenever evaluation can be separated from implementation; Appendix~\ref{sec:app_examples} gives regression, structural VAR, and inflation-forecasting sketches, and Appendix~\ref{sec:app_formal} provides a more technical exposition of the search problem.}

\subsection{Running the loop, and the holdout rule}
\label{sec:running}

\rev{No custom software is required. The researcher opens a general-purpose coding assistant, such as Claude Code, Codex, or OpenCode, and asks it to implement the instruction contract. This step can be as simple as prompting ``Read program.md and implement.'' The agent then reads the contract, edits the allowed script, runs the fixed evaluator, logs the result, and repeats the loop in Algorithm~\ref{alg:loop} autonomously until the budget is exhausted.}

\rev{After the search, the selected script can be evaluated on a post-search holdout $D^{H}$. The holdout rule is simple: $D^{H}$ is scored once, by the researcher, after the search has ended. It is not used by the evaluator and is not available in the agent's workspace. In our application, the holdout was assembled only after the agentic search was complete, so no holdout data existed in the working directory during the search.}

\subsection{Scope and caveats}
\label{sec:caveats}

\rev{Several caveats bound what the protocol can promise. First, residual look-ahead remains possible: a language model may encode background knowledge overlapping the holdout period, so a holdout clean in the workspace need not be clean in the model's memory. Holdout results are therefore a guardrail rather than proof. Second, the protocol disciplines search, not inference; it is not a substitute for formal post-selection inference. Third, the implementation is illustrative rather than optimal. The greedy keep-or-discard loop could be replaced by evolutionary or tree search \citep{NovikovEtAl2025,AygunEtAl2025}.}

\rev{Finally, the protocol does not guarantee semantic compliance. Written instructions do not bind an agent the way executable code does: the loop rewards scripts that improve the evaluator's score, while many substantive restrictions in the contract are stated in natural language. For example, instructions such as ``do not include higher order terms'' or ``do not combine with other methods'' are semantic, and the agent may trade them off against score improvements in an opaque way. The search can therefore drift from the researcher's intended model class. This is why auditability is essential: the protocol records the search path, so such drift is visible rather than hidden.}

\section{Empirical illustration: real-time tuning of peLASSO}
\label{sec:empirical}

\subsection{The tuning problem in Diebold and Shin (2019)}
\label{sec:problem}

\rev{We consider a forecast-combination problem. There are $K$ forecasts for the same target, and the goal is to construct a combined forecast, a function of these forecasts, with lower out-of-sample RMSE. In this setting, the simple average, which assigns equal weights in a linear aggregation, is famously hard to beat with estimated weights. One explanation is that estimation uncertainty can dominate the gains from estimating combination weights. This motivates regularized combination methods.}

\rev{\citet{DieboldShin2019}'s partially egalitarian LASSO (peLASSO) selects forecasters using LASSO, then shrinks the surviving weights toward equality. Implemented in real time, it requires choosing two regularization parameters at each forecast origin. The original paper instead reports ex post optimal parameters and proposes easier-to-tune subset averaging rules inspired by the peLASSO analysis, leaving no data-based rule for real-time peLASSO tuning. On their euro area GDP data, ex post peLASSO attains RMSE 1.400 versus 1.504 for the simple average, a seven percent oracle gap. The present application takes this unresolved tuning problem as the object of agentic search.}

\subsection{Search design and samples}
\label{sec:design}

\rev{We keep this out-of-sample evaluation setting. The target is year-on-year euro area real GDP growth, and the inputs are $K = 23$ individual SPF forecasts over the 1999Q3--2016Q4 sample. The score is RMSE over 65 forecast dates, 2000Q4--2016Q4, matching the study's burn-in convention.}

\rev{Our starting point is the exact codebase that reproduces the study. We make two modifications to implement the protocol of Section~\ref{sec:protocol}:}
\begin{enumerate}
\item \rev{\textit{Score function (\texttt{prepare.R}).} This function calculates out-of-sample RMSE for a given estimation function, \texttt{train.R}. If \texttt{train.R} takes the simple average of the current forecasts, the score function exactly reproduces the original simple-average benchmark.}
\item \rev{\textit{Instruction contract (\texttt{program.md}).} We write the contract to instruct the agent: ``Find the best look-ahead-free, data-based lambda selection algorithm for PE Lasso forecast combination.''\footnote{\rev{The full instruction contract is about 835 words because it also gives project context, describes the replicated study and directory structure, states the edit and no-look-ahead rules, and specifies logging and recovery instructions. Appendix~\ref{sec:app_program} reproduces the exact verbatim contract.}}}
\end{enumerate}
\rev{Together with \texttt{train.R} as the editable script and \texttt{results.tsv} as the experiment log, these files implement the four objects in Table~\ref{tab:files}. We run three independent searches, totaling 661 experiments. As benchmarks, we retain the simple average, ex post peLASSO, and the two strongest subset averaging methods in that paper: best $\leq 6$-average and best $(\leq 6, \leq 40)$-average (Tables~4 and~5).}

\rev{After the search, we perform holdout evaluation. The split follows naturally from the replication boundary: that same codebase contains data only through 2016Q4 and uses the same out-of-sample rule. We extended the series to 2017Q1--2025Q4 (36 quarters) only after the agentic search ended, so these later realizations form a natural post-search holdout.}

\subsection{Empirical findings}
\label{sec:results}

\paragraph{Final scores.} Table~\ref{tab:summary} reports RMSE for the search sample and the holdout. On the search sample, all three runs beat both the simple average and the ex post peLASSO benchmark, with relative RMSEs of 0.858, 0.510, and 0.808 for Runs~1--3.

\begin{table}[t!]
\centering
\begin{threeparttable}
\caption{Search-sample and holdout evaluation}
\label{tab:summary}
\footnotesize
\begin{tabular}{@{}l S[table-format=1.3] S[table-format=1.3] S[table-format=1.3] S[table-format=1.3] S[table-format=1.3] S[table-format=1.3]@{}}
\toprule
 & \multicolumn{2}{c}{Search sample} & \multicolumn{2}{c}{Holdout} & \multicolumn{2}{c}{Holdout excl.\ COVID} \\
 \cmidrule(lr){2-3} \cmidrule(lr){4-5} \cmidrule(lr){6-7}
 & {RMSE} & {Relative} & {RMSE} & {Relative} & {RMSE} & {Relative} \\
\midrule
Simple average          & 1.504 & 1.000 & 2.979 & 1.000 & 1.120 & 1.000 \\
peLASSO ex post          & 1.400 & 0.930 & {2.964\,[.191]} & 0.995 & {1.075\,[.136]} & 0.960 \\
Best $\leq 6$-avg           & 1.435 & 0.954 & {2.901\,[.240]} & 0.974 & {1.043\,[.347]} & 0.932 \\
Best $(\leq 6, \leq 40)$-avg        & 1.378 & 0.916 & {2.901\,[.127]} & 0.974 & {1.142\,[.602]} & 1.020 \\
\midrule
Run 1 (232 experiments)                  & 1.291 & 0.858 & {2.816\,[.177]} & 0.945 & {0.965\,[.234]} & 0.861 \\
Run 2 (228 experiments)                  & 0.767 & 0.510 & {2.417\,[.127]} & 0.811 & {0.827\,[.108]} & 0.739 \\
Run 3 (201 experiments)                  & 1.216 & 0.808 & {3.244\,[.852]} & 1.089 & {1.305\,[.755]} & 1.165 \\
\bottomrule
\end{tabular}
\begin{tablenotes}\small
\item \textit{Notes:} RMSE of forecast errors for euro area real GDP growth (year on year). \rev{Search sample: 1999Q3--2016Q4 data; RMSE scored over 65 forecast dates (2000Q4--2016Q4).} Holdout: 2017Q1--2025Q4 (36~quarters). Holdout excl.\ COVID: holdout dropping 2020Q1--Q4 (32~quarters). Relative RMSE is computed against the simple average on each respective sample. Best $\leq 6$-avg and best $(\leq 6, \leq 40)$-avg are from the original paper, Tables~4 and~5. Bracketed values in the holdout columns are one sided $p$-values from the Diebold--Mariano test that the method has superior predictive accuracy relative to the simple average, using the EWC fixed-$b$ approximation \citep{ShinSchor2026}; these have intrinsically low power given the small holdout sample and serial correlation in forecast errors, and should be read as descriptive. peLASSO ex post re-optimizes $\lambda$ on each evaluation window separately; see Appendix~\ref{sec:app_holdout} for the fixed-$\lambda$ variant. Full results including intermediate methods are in Appendix~\ref{sec:app_holdout}.
\end{tablenotes}
\end{threeparttable}
\end{table}

On the holdout, Run~2 remains strongest, with relative RMSE 0.811 (0.739 excluding the COVID quarters of 2020), and Run~1 also holds up, at 0.945 (0.861 excluding COVID). The two benchmark methods from the original paper generalize modestly, each at 0.974. Thus, even on the holdout, Runs~1 and~2 outperform every non-agent benchmark in Table~\ref{tab:summary}, including the best methods from the original paper. In contrast, Run~3 (1.089) performs worse than the simple average on the holdout, and peLASSO ex post shows essentially no improvement (0.995). Agentic search can uncover methods that generalize, but also sample-specific improvements.

\paragraph{What the log reveals.} \rev{The log shows how the agent moves from tuning to method discovery.\footnote{\rev{The appendices archive the full instruction contract, a detailed log excerpt, method details, and additional holdout and robustness checks.}} Each run begins with the same tension: the contract says ``Do not try to combine with other methods,'' but also encourages ``combining previous near-misses'' and ``more radical approaches'' if progress stalls. Early attempts stay close to peLASSO tuning, trying cross-validation and related lambda-selection rules. As gains become harder, the paths diverge. Run~1 moves toward stability selection and performance weighting. Run~2 stops treating lambda choice as the only margin and builds a ranking-based combination with bias correction. Run~3 stays closest to peLASSO tuning, using adaptive LASSO with forward cross validation, but eventually adds an egalitarian elastic-net blend. The useful information is not only that drift occurred, but when it occurred and in what direction. Final scores alone would hide this path; the log makes the tradeoff between semantic instructions and score improvement visible.}

\paragraph{What we learn.} \rev{Agentic search is powerful and useful for facilitating empirical research by automating specification search at scale. In our experiment, it found methods that beat standard benchmarks and, in two runs, survived a post-search holdout. It is also adaptive search over many programs, so search-sample improvements can overfit and semantic instructions can be traded off against score improvements. The lesson is not to trust the final script alone. The useful object is the auditable bundle: fixed evaluator, confined edit surface, complete log, and a holdout scored once after the search.}

\bigskip
\section*{Declaration of generative AI and AI-assisted technologies in the manuscript preparation process}
\rev{During the preparation of this work, the author used Claude Opus~4.6 and GPT~5.4 for editorial and programming assistance, including code cleanup and revision of parts of the manuscript draft; the agentic search experiments themselves used Claude Code as described in the paper. The author reviewed and edited the output as needed and takes full responsibility for the content of the published article.}

\bibliographystyle{plainnat}
\bibliography{references}

%
%
\newpage

\begin{center}
{\LARGE \textbf{Online Appendix}} \\[12pt]
{\large An Auditable AI Agent Loop for Empirical Economics} \\[12pt]
{\LARGE \textbf{\textit{Not for publication}}}
\end{center}

\bigskip

\appendix

\section{Implementation details}
\label{sec:app_implementation}

\subsection{Four-file architecture}

The protocol requires only four files in a shared directory:
\[
C \equiv \texttt{program.md}, \quad
\tau \equiv \texttt{train.R}, \quad
S \equiv \texttt{prepare.R}, \quad
L \equiv \texttt{results.tsv}.
\]
\rev{Section~\ref{sec:fourfiles} of the main text describes the role of each file; this appendix records the implementation details.} \texttt{prepare.R} loads the search sample, sources \texttt{train.R}, and returns the scalar score; the holdout data is entirely outside the agent's workspace. \texttt{results.tsv} is a tab-separated audit log in which each row records a commit identifier, the score, the outcome status, and a verbal description of the strategy attempted. In the working directory, the git commit history preserves the actual code of every retained~$\tau_k$.\footnote{The replication archive includes the TSV logs and final code but not the full git object store. Full recoverability from commit history requires access to the original working directory.}

The four-file architecture is adapted from \citeauthor{Karpathy2026}'s (\citeyear{Karpathy2026}) \texttt{autoresearch} repository.\footnote{See \url{https://github.com/karpathy/autoresearch}. The architecture is representative of a broader class of agent loops in which a language model acts as the optimizer within a fixed evaluation harness.} The contribution here is not the loop itself but its translation into a protocol for empirical economics: the immutable evaluator fixes what counts as success, the editable script defines the search surface, the instruction contract specifies the researcher's permissions, and the audit log turns every attempted specification into a reportable object.

\subsection{Triggering the agent loop}

In the simplest case, the researcher opens a general-purpose coding assistant and prompts it to implement \texttt{program.md}. The agent reads the instruction contract, which describes the experiment loop of Algorithm~\ref{alg:loop} in plain language, and begins the edit--evaluate--log--retain/revert cycle autonomously.

In our application, we used an outer shell script (\texttt{run.sh}) to restart agent sessions when the model reached its context window limit. Each restart reads \texttt{results.tsv} and the git history to recover state, and the session prompt injects the remaining experiment budget: ``You have a budget of $K_{\mathrm{remaining}}$ more experiments. Stop after that.'' This is how the budget $K$ is communicated and enforced in practice. The outer script also provides an approximate stop by counting rows in \texttt{results.tsv} and refusing to launch a new session once the count reaches $K$; a session already in progress may overshoot. The full \texttt{run.sh} is available in the replication archive.

\subsection{Isolating the holdout}

The post-search holdout $D^{H}$ must remain entirely outside the agent's workspace during the search. In our application, the holdout period (2017Q1--2025Q4) was constructed after the agentic search was complete, so no holdout data existed in the working directory while the agent was running. More generally, researchers should ensure that holdout data is not accessible to the agent during the search. Because current coding agents can access the internet and execute arbitrary code, an agent could in principle download or construct holdout period data on its own. Disabling internet access during the search, restricting the agent's file system permissions, or simply withholding the holdout data from the working directory are practical safeguards. Without such boundaries, the transparency of the holdout evaluation cannot be guaranteed.

\subsection{Actual run}

Each run uses Claude Code with Opus 4.6 at default thinking effort, starting from a fresh agent with the same initial script (simple average) and an approximate budget of $K = 200$ experiments. The evaluator calls the forecasting function 66 times ($t = 5, \ldots, 70$) but scores RMSE over 65 periods ($t = 6, \ldots, 70$), matching the burn-in and evaluation design in \citet{DieboldShin2019}. Table~\ref{tab:experiments} reports the experiment counts for the three agent runs. The budget is enforced as a soft constraint; the realized counts are 232, 228, and 201 for Runs~1, 2, and~3, respectively. If a candidate script exceeds 30 minutes of runtime, the experiment is terminated and logged as a crash.

\begin{table}[t!]
\centering
\begin{threeparttable}
\caption{Experiment counts}
\label{tab:experiments}
\small
\begin{tabular}{@{}l cccc@{}}
\toprule
 & {Experiments} & {Keeps} & {Discards} & {Crashes} \\
\midrule
Run 1 & 232 & 42 & 188 & 2 \\
Run 2 & 228 & 61 & 164 & 3 \\
Run 3 & 201 & 46 & 152 & 3 \\
\bottomrule
\end{tabular}
\begin{tablenotes}\small
\item \textit{Notes:} ``Keep'' means the modification improved on the current best RMSE; ``discard'' means it did not; ``crash'' means the candidate script exceeded the 30-minute timeout or produced an error. The budget is enforced as a soft constraint; realized counts vary.
\end{tablenotes}
\end{threeparttable}
\end{table}

\section{Instruction contract}
\label{sec:app_program}

The following is the verbatim \texttt{program.md} used in the empirical application of Section~\ref{sec:empirical}.

{\scriptsize
\begin{verbatim}
# autoresearch — PE Lasso lambda selection

## Goal

Find the best look-ahead-free, data-based lambda selection algorithm
for PE Lasso forecast combination.

PE Lasso forecast combination we mean partially egalitary lasso method
in the paper /original_paper/DiebodShinEgalitarianLasso.md
(both one-step conceptualization and two-step implementation).
For both cases, we have two regularization parameters.

IMPORTANT: Focus on two-step implementation of PE Lasso. Do not try
to combine with other methods. I want to focus on finding a way to
choose these regularization parameters for PE Lasso so that it becomes
tuning free method. We could try bunch of different CV methods. Or,
other data adaptive method automatically selects these tuning
parameters. We understand it is difficult problem finding two
regularization parameters from two different stages. But, this is
exactly why we want to run this experiments.

The simple average benchmark has RMSE ~1.50. The ex-post oracle
(which cheats by using future data) achieves ~1.40. Your job is to
close that gap.

You decide what strategies to explore. The reference implementation
is in `original_program/R/` — read it for ideas.

## Setup

1. Agree on a run tag with the user: propose a tag based on today's
   date (e.g. `mar11`). The branch `autoresearch/<tag>` must not
   already exist.
2. Create the branch: `git checkout -b autoresearch/<tag>`
3. Read the in-scope files:
   - `program.md` — this file (your instructions)
   - `prepare.R` — fixed evaluation harness. Do not modify.
   - `train.R` — the file you modify. Defines `select_lambda(info)`.
4. Check session history: Read `results.tsv` and `git log` to
   understand what has been tried before. This is your memory across
   sessions.
5. Initialize results.tsv if it doesn't exist: create it with just
   the header row.
6. Confirm and go.

## The rules

- One file: You only edit `train.R`. It must define
  `select_lambda(info)` returning `list(forecast, method)`.
- No lookahead: At time t, only data from 1:(t-1) is available.
  The `info` object enforces this.
- No new packages: Only glmnet, forecast, and base R.
- Runtime: Each full run (66 evaluation periods) should complete
  in under 30 minutes.
- Deterministic: Use a fixed seed if randomness is needed.

## What's in `info`

  info$X_train      # rolling window training forecasts [max(1,t-w):(t-1), 23]
  info$y_train      # corresponding actuals
  info$X_history    # full history [1:(t-1), 23]
  info$y_history    # full history actuals
  info$x_new        # current period forecasts [23]
  info$lambda_grid  # 200-point lambda grid, exp(15) to exp(-15)
  info$lambda_grid2 # same grid (for two-step methods)
  info$t            # current time index
  info$w            # window size (20)

All functions from `prepare.R` are in scope: `pelasso_lasso_avg()`,
`pelasso_lasso_eridge()`, `pelasso_lasso_elasso()`, `fit_lasso()`,
`fit_ridge()`, `fit_elasso()`, `fit_eridge()`,
`cv_loo_bandwidth()`, `cv_loo_bandwidth_2step()`, etc.

## Benchmarks

| Method                               | RMSE  | Notes                      |
|--------------------------------------|-------|----------------------------|
| Simple average                       | ~1.50 | Baseline (no lambda)       |
| Ex-post optimal peLASSO(LASSO, Avg)  | ~1.40 | Oracle — uses future data  |
| Good data-based method               | 1.42-1.48 | Realistic target range |

## Output format

The script prints:

  ---
  method:         <name>
  rmse:           <value>
  benchmark_rmse: <simple average RMSE>
  relative_rmse:  <ratio>

Extract the key metric: `grep "^rmse:" run.log`

## Logging results

Log every experiment to `results.tsv` (tab-separated).
Do NOT commit this file.

  commit   rmse    status   description
  a1b2c3d  1.5012  keep     baseline simple average
  b2c3d4e  1.4723  keep     LOO CV for pelasso_lasso_avg
  c3d4e5f  1.4891  discard  BIC-based lambda selection
  d4e5f6g  0.0000  crash    2D CV OOM on full grid

## Session recovery

You may be starting a fresh session after a previous one was
interrupted (context limit, crash, etc.). On every session start:

1. Read `results.tsv` and `git log --oneline` to see what has
   been tried.
2. Reconcile state: Compare the last `results.tsv` entry to
   the latest git commit.
   - If the latest commit is NOT in `results.tsv`, it was never
     evaluated. Run it now and log the result before moving on.
   - If `train.R` has uncommitted changes, they are leftover from
     an interrupted edit. Review and either commit or discard.
3. Identify the current best RMSE from `results.tsv` (the lowest
   `rmse` with status `keep`).
4. Continue experimenting from there.

## The experiment loop

LOOP (until experiment budget is exhausted or interrupted):

1.  Read the current state: `train.R`, `results.tsv`, git log
2.  Edit `train.R` with a new lambda selection strategy
3.  `git commit -m "descriptive message"`
4.  Run: `Rscript prepare.R > run.log 2>&1`
5.  Extract: `grep "^rmse:" run.log`
6.  If empty (crash): `tail -n 50 run.log` to diagnose
7.  Log to `results.tsv`
8.  If RMSE improved -> keep the commit
9.  If RMSE worse -> `git reset --hard HEAD~1`
10. Repeat

Logging is step 7, BEFORE the keep/revert decision in steps 8-9.
This ensures every experiment is recorded even if the session is
interrupted right after.

Timeout: If a run exceeds 30 minutes, kill it and treat as crash.

Crashes: Fix trivial bugs (typos, off-by-one). If the idea is
fundamentally broken, log crash and move on.

NEVER STOP: Do not pause to ask if you should continue. Run
autonomously until your experiment budget runs out or you are
manually interrupted. If you run out of ideas, re-read
`original_program/R/` for new angles, try combining previous
near-misses, or try more radical approaches.
\end{verbatim}
}

\section{Experiment log excerpt}
\label{sec:app_log}

Table~\ref{tab:log} presents a selected excerpt from the 201-experiment log of Run~3. The trajectory illustrates a systematic exploration pattern. The agent begins with standard cross-validation for peLASSO (experiments~1--4), quickly moves to adaptive LASSO with ridge-based penalty weights and forward cross-validation (experiments~25--45), discovers that multi-horizon forward CV improves regularization parameter selection (experiment~86), introduces model averaging over near-optimal lambda values (experiment~113), and finally arrives at an ensemble that blends peLASSO with an egalitarian elastic net component (experiments~138--201). Each phase builds on the structural lessons of the previous one.

\begin{table}[t!]
\centering
\begin{threeparttable}
\caption{Experiment log excerpt: Run~3}
\label{tab:log}
\small
\begin{tabular}{@{}cl S[table-format=1.4] l@{}}
\toprule
{Exp.} & {Status} & {RMSE} & {Description} \\
\midrule
1   & keep    & 1.5043 & baseline simple average \\
2   & keep    & 1.4692 & pelasso(lasso,avg) LOO CV B=0 \\
3   & crash   & {---}  & pelasso(lasso,avg) LOO CV B=1 -- zero variance error \\
4   & discard & 1.9556 & pelasso(lasso,eridge) 2D LOO CV coarse 30pt grid \\
25  & keep    & 1.4455 & pe(adalasso,avg) fwdCV exp 0.95 ridge penalty weights \\
45  & keep    & 1.3839 & pe(adalasso,avg) rolling window ridge penalties decay=0.6 \\
86  & keep    & 1.3539 & pe(adalasso,avg) h=2 step ahead forward CV LOO ridge \\
113 & keep    & 1.3202 & pe(adalasso,avg) h=2 model averaging top 2pct lambdas \\
138 & keep    & 1.2533 & pe(adalasso,blend) 80pct PE + 20pct eridge ensemble \\
152 & keep    & 1.2253 & pe(adalasso,blend) 70/30 PE + eEN cv.glmnet LOO \\
201 & keep    & 1.2159 & pe(adalasso,blend) model avg 1pct h1(0.05)+h2(0.95) \\
\bottomrule
\end{tabular}
\begin{tablenotes}\small
\item \textit{Notes:} Selected rows from the 201-experiment log of Run~3. Descriptions are the agent's verbatim one-line summaries, truncated for display. Experiment~1 is the initial simple average baseline. RMSE for crash entries is undefined; the implementation records them as zero in \texttt{results.tsv}, but they are excluded from all comparisons (formally, $s_k = +\infty$). Full logs are in the replication archive.
\end{tablenotes}
\end{threeparttable}
\end{table}

\section{Discovered methods}
\label{sec:app_methods}

Each agent run produced a final method and several intermediate variants. The holdout evaluation in Section~\ref{sec:app_holdout} reports results for the final method and two representative intermediates per run, selected from distinct algorithmic phases. We describe each method here.

\rev{As stated in Section~\ref{sec:design}, all} methods operate within the same rolling window framework \rev{except where noted}. At each forecast origin $t$, let $\mathbf{X}_t \in \mathbb{R}^{n_t \times K}$ and $\mathbf{y}_t \in \mathbb{R}^{n_t}$ denote the training window of forecaster predictions and GDP realizations, where $n_t = \min(t - 1, W)$, $W = 20$, and $K = 23$. Let $\mathbf{x}_t = (x_{1,t}, \ldots, x_{K,t})'$ denote the new forecaster predictions for period $t$. Each method produces a combined forecast $\hat{y}_t$; most methods use only $(\mathbf{X}_t, \mathbf{y}_t, \mathbf{x}_t)$, with exceptions noted below.

\subsection{Run~1: stability selection with performance weighting}

\paragraph{Run~1 (final).} Stability selection with performance weighted aggregation.
\begin{enumerate}
\item Fit elastic net ($\alpha = 0.65$) with 5 fold CV on the training window. Define $\hat{\lambda}_{1.5\mathrm{se}}$ as the largest $\lambda$ such that the CV error is within $1.5$ standard errors of the minimum.
\item \textit{Stability selection.} Let $\mathcal{W} = \{j : |\log\lambda_j - \log\hat{\lambda}_{1.5\mathrm{se}}| < \delta\}$ with $\delta = 0.25$. The selection frequency of forecaster $k$ is
\begin{equation}
\pi_k = |\mathcal{W}|^{-1} \sum_{j \in \mathcal{W}} \mathbf{1}\{\hat{\beta}_k(\lambda_j) \neq 0\}.
\end{equation}
The active set is $\mathcal{A} = \{k : \pi_k > 0.4\}$.
\item \textit{Composite weights.} Using the $m = 2$ most recent training observations, compute individual $\mathrm{RMSE}_k$ for each $k \in \mathcal{A}$ and form
\begin{equation}
w_k = \pi_k^2 \cdot \mathrm{RMSE}_k^{-p}, \quad p = 14.
\end{equation}
The exponent $p = 14$ was selected by the search from nearby alternatives (powers 2 through 20 were tried).
\item \textit{Dominance check.} If $w_{(1)} > 5 \cdot w_{(2)}$ (where $w_{(j)}$ denotes the $j$th largest weight), use the single best forecaster.
\item Otherwise, the combined forecast is the weighted $q$th quantile ($q = 0.44$) of $\{x_{k,t}\}_{k \in \mathcal{A}}$ with normalized weights $w_k / \sum_k w_k$, computed by linear interpolation of the weighted empirical CDF.
\end{enumerate}

\paragraph{Run~1a (Phase~2: elastic net with median).} PE selection using elastic net ($\alpha = 0.65$) with 5 fold CV and $\hat{\lambda}_{1\mathrm{se}}$. The combined forecast is the median of the selected forecasters:
\begin{equation}
\hat{y}_t = \mathrm{median}\bigl(\{x_{k,t}\}_{k \in \mathcal{A}(\hat{\lambda}_{1\mathrm{se}})}\bigr).
\end{equation}
No stability selection, no performance weighting.

\paragraph{Run~1b (Phase~1: peLASSO with CV).} Standard peLASSO(LASSO, Avg) with $\hat{\lambda}_{1\mathrm{se}}$ from \texttt{cv.glmnet}:
\begin{equation}
\hat{y}_t = |\mathcal{A}(\hat{\lambda}_{1\mathrm{se}})|^{-1} \sum_{k \in \mathcal{A}(\hat{\lambda}_{1\mathrm{se}})} x_{k,t}.
\end{equation}
This is the closest variant to the original \citet{DieboldShin2019} method, but with data driven $\lambda$ selection rather than ex post optimization.

\subsection{Run~2: ranking, adaptive windows, and bias correction}

\paragraph{Run~2 (final).} Temporally weighted ranking with adaptive window selection and bias correction.

For a given training window $(\mathbf{X}, \mathbf{y})$ of size $n$:
\begin{enumerate}
\item \textit{Temporal weights.} Define $\omega_s = \exp\bigl(-\alpha(n - s)/(n - 1)\bigr) / \sum_{j=1}^n \exp\bigl(-\alpha(n - j)/(n - 1)\bigr)$ with $\alpha = 6$. More recent observations receive higher weight.
\item \textit{Weighted MAE ranking.} $\mathrm{MAE}_k = n^{-1}\sum_{s=1}^n n\omega_s |y_s - x_{k,s}|$. Rank forecasters by ascending MAE.
\item \textit{LOO CV for subset size.} For each candidate $N \in \{3, \ldots, \min(K, 18)\}$ and each left out observation $i$:
\begin{itemize}
\item Rerank by RMSE on the remaining $n - 1$ observations.
\item Select the top $N$ forecasters, weight them by $\tilde{w}_j \propto \mathrm{MAE}_j^{-3}$ (on leave out data).
\item Compute the weighted average LOO forecast $\hat{y}_i^{\mathrm{LOO}} = \sum_j \tilde{w}_j x_{j,i}$ and the median based LOO forecast $\hat{y}_i^{\mathrm{pct}} = Q_{0.50}(\{x_{j,i}\}_{j \in \text{top } N})$.
\end{itemize}
Select $N^* = \arg\min_N \sqrt{n^{-1}\sum_i (\hat{y}_i^{\mathrm{LOO}} - y_i)^2}$. After $N^*$ is selected, the current period forecast (step~5) uses the full window weighted MAE ranking, not the LOO reranking or the $\mathrm{MAE}^{-3}$ weighted average.
\item \textit{Adaptive window.} Evaluate steps 1--3 at the full rolling window and at each sub-window $w' \in \{4, 5, \ldots, 19\}$, keeping whichever $(N^*, w')$ combination yields the lowest LOO RMSE.
\item \textit{Median aggregation.} The base forecast is $\tilde{y}_t = Q_{0.50}(\{x_{k,t}\}_{k \in \text{top } N^*})$.
\item \textit{Bias correction.} The final forecast applies a bias adjustment based on the most recent leave one out error:
\begin{equation}\label{eq:bias_correction}
\hat{y}_t = \tilde{y}_t - \gamma \cdot \hat{e}_n^{\mathrm{LOO,pct}}, \quad \gamma = 0.80,
\end{equation}
where $\hat{e}_n^{\mathrm{LOO,pct}} = \hat{y}_n^{\mathrm{pct}} - y_n$ is the percentile based LOO error from the most recent training observation.
\end{enumerate}

\paragraph{Run~2a (no bias correction).} Identical to Run~2 (steps 1--5) but omits the bias correction~\eqref{eq:bias_correction}:
\begin{equation}
\hat{y}_t = \tilde{y}_t.
\end{equation}
This isolates the contribution of the forecaster ranking and adaptive window mechanism.

\paragraph{Run~2b (weighted ranking, no adaptive window).} Steps 1--2 as in Run~2 with a fixed window. Step 3 uses a penalized LOO criterion, $\mathrm{RMSE}_{\mathrm{LOO}}(N) + 0.01 \cdot \log(N + 1)$, to select $N^*$. The combined forecast is a weighted average (not median) of the top $N^*$ forecasters:
\begin{equation}
\hat{y}_t = \sum_{j=1}^{N^*} \tilde{w}_j \, x_{\sigma(j),t}, \quad \tilde{w}_j \propto \mathrm{MAE}_{\sigma(j)}^{-3},
\end{equation}
where $\sigma$ denotes the MAE ranking. No adaptive window, no bias correction.

\subsection{Run~3: adaptive LASSO with forward CV and ensemble}

\paragraph{Run~3 (final).} Adaptive LASSO with multi horizon forward CV, model averaging, and egalitarian elastic net ensemble.
\begin{enumerate}
\item \textit{Pilot estimator.} Fit elastic net ($\alpha_{\mathrm{pilot}} = 0.1$) with LOO CV on the rolling window to obtain $\hat{\boldsymbol{\beta}}^{\mathrm{ridge}}$.
\item \textit{Adaptive LASSO penalties.} Define forecaster specific penalty factors
\begin{equation}
\tilde{w}_k = \frac{(|\hat{\beta}_k^{\mathrm{ridge}}| + \epsilon)^{-1}}{\overline{(|\hat{\beta}^{\mathrm{ridge}}| + \epsilon)^{-1}}}, \quad \epsilon = 5 \times 10^{-3},
\end{equation}
where $\overline{\cdot}$ denotes the cross sectional mean. Fit LASSO with penalty $\lambda \cdot \tilde{w}_k |\beta_k|$ on the full history.
\item \textit{Multi horizon forward CV.} Lambda is selected by walk forward CV on the full history preceding the current origin, rows $1, \ldots, t - 1$. For fold $f$, fit adaptive LASSO on the first $t_f$ rows and compute forecast errors on the next two rows $t_f + 1$ and $t_f + 2$ (where $t_f + 2 \leq t - 1$ by construction). Let $e_{f,s}(\lambda) = \hat{y}_{t_f+s}(\lambda) - y_{t_f+s}$ for $s = 1, 2$, where $\hat{y}_{t_f+s}(\lambda)$ applies the $\lambda$ dependent combination rule to the forecaster predictions at row $t_f + s$. Two candidate aggregation rules are evaluated at each $\lambda$: simple average and blended average ($0.7 \times \text{mean} + 0.3 \times \text{ridge weighted mean}$). The CV criterion is
\begin{equation}
\mathrm{CV}(\lambda) = \sum_{f} w_f \left(0.05 \cdot e_{f,1}^2(\lambda) + 0.95 \cdot e_{f,2}^2(\lambda)\right),
\end{equation}
with fold weights $w_f \propto \gamma^{F - f}$ and $\gamma = 0.75$.
\item \textit{Model averaging.} Average forecasts over all $\lambda$ within 1\% of the best CV score, weighted by inverse RMSE.
\item \textit{Egalitarian elastic net.} Fit elastic net ($\alpha = 0.5$) via \texttt{cv.glmnet} with LOO CV on the rolling window, selecting $\hat{\lambda}_{\min}$, on the egalitarian transformation $\tilde{y}_s = y_s - K^{-1}\sum_k x_{k,s}$, obtaining $\tilde{\boldsymbol{\beta}}^{\mathrm{eEN}}$, and compute $\hat{y}_t^{\mathrm{eEN}} = \sum_k (\tilde{\beta}_k^{\mathrm{eEN}} + K^{-1})\, x_{k,t}$.
\item \textit{Final blend.}
\begin{equation}\label{eq:run07_blend}
\hat{y}_t = 0.70 \cdot \hat{y}_t^{\mathrm{PE}} + 0.30 \cdot \hat{y}_t^{\mathrm{eEN}}.
\end{equation}
\end{enumerate}

\paragraph{Run~3a (Phase~4: single horizon, no blend).} Steps 1--2 as in Run~3. Forward CV evaluating only at the next panel row ($t_f + 1$) with $\gamma = 0.75$. No model averaging. Simple average of selected forecasters. No egalitarian elastic net blend:
\begin{equation}
\hat{y}_t = |\mathcal{A}(\hat{\lambda})|^{-1} \sum_{k \in \mathcal{A}(\hat{\lambda})} x_{k,t}.
\end{equation}

\paragraph{Run~3b (Phase~5: multi horizon, no blend).} Same pilot and adaptive penalty construction as Run~3, with $\gamma = 0.80$ and $\epsilon = 10^{-3}$. Forward CV evaluates only the simple average aggregation rule (no blended alternative). Model averaging over $\lambda$ values within 1\% of the best CV score is applied. No egalitarian elastic net blend: $\hat{y}_t = \hat{y}_t^{\mathrm{PE}}$.

\section{Post-search holdout evaluation}
\label{sec:app_holdout}

This section reports the full holdout evaluation summarized in Section~\ref{sec:empirical} of the main text. The holdout period (2017Q1--2025Q4, 36~quarters) was withheld from the evaluator during the agentic search.

\subsection{Data and processing}

The extended dataset combines ECB Survey of Professional Forecasters (SPF) individual microdata with Eurostat GDP realizations (table \texttt{namq\_10\_gdp}, chain linked volumes, seasonally and calendar adjusted). The original sample (1999Q3--2016Q4, 70~observations) is copied verbatim from the original \texttt{H1\_gdp.csv} of \citet{DieboldShin2019} to guarantee exact RMSE reproduction. The new sample (2017Q1--2025Q4, 36~quarters) is constructed from the SPF extraction. Missing forecaster responses in the new sample are imputed by linear interpolation for interior gaps and by cross-sectional mean for trailing gaps and inactive forecasters. COVID quarters are defined as 2020Q1--2020Q4.

\subsection{Full holdout results}

Table~\ref{tab:holdout_full} reports RMSE for all methods, including intermediate variants that isolate individual components. For each run, we include two representative intermediate methods selected from distinct algorithmic phases. The intermediate methods are informative because they isolate the contribution of individual components such as stability selection, bias correction, or ensemble blending.

\begin{table}[htbp]
\centering
\begin{threeparttable}
\caption{Full holdout evaluation of forecast combination methods}
\label{tab:holdout_full}
\footnotesize
\begin{tabular}{@{}l S[table-format=1.3] S[table-format=1.3] S[table-format=1.3] S[table-format=1.3] S[table-format=1.3] S[table-format=1.3]@{}}
\toprule
 & \multicolumn{2}{c}{Search sample} & \multicolumn{2}{c}{Holdout} & \multicolumn{2}{c}{Holdout excl.\ COVID} \\
 \cmidrule(lr){2-3} \cmidrule(lr){4-5} \cmidrule(lr){6-7}
 & {RMSE} & {Relative} & {RMSE} & {Relative} & {RMSE} & {Relative} \\
\midrule
\multicolumn{7}{@{}l}{\textit{Panel A: Benchmarks and original paper methods}} \\[3pt]
Simple average          & 1.504 & 1.000 & 2.979 & 1.000 & 1.120 & 1.000 \\
peLASSO ex post (original)          & 1.400 & 0.930 & {3.172\,[.880]} & 1.065 & {1.360\,[.885]} & 1.214 \\
peLASSO ex post (per window)   & 1.400 & 0.930 & {2.964\,[.191]} & 0.995 & {1.075\,[.136]} & 0.960 \\
Best $\leq 6$-avg           & 1.435 & 0.954 & {2.901\,[.240]} & 0.974 & {1.043\,[.347]} & 0.932 \\
Best $(\leq 6, \leq 40)$-avg        & 1.378 & 0.916 & {2.901\,[.127]} & 0.974 & {1.142\,[.602]} & 1.020 \\
Best individual         & 1.403 & 0.933 & {2.994\,[.716]} & 1.005 & {1.164\,[.714]} & 1.040 \\
\midrule
\multicolumn{7}{@{}l}{\textit{Panel B: Run~1 family}} \\[3pt]
Run 1 (final)                       & 1.291 & 0.858 & {2.816\,[.177]} & 0.945 & {0.965\,[.234]} & 0.861 \\
\quad Run 1a (ENet, median)         & 1.390 & 0.924 & {2.991\,[.742]} & 1.004 & {1.158\,[.794]} & 1.034 \\
\quad Run 1b (peLASSO, CV)          & 1.451 & 0.964 & {3.059\,[.938]} & 1.027 & {1.270\,[.929]} & 1.134 \\
\midrule
\multicolumn{7}{@{}l}{\textit{Panel C: Run~2 family}} \\[3pt]
Run 2 (final)                       & 0.767 & 0.510 & {2.417\,[.127]} & 0.811 & {0.827\,[.108]} & 0.739 \\
\quad Run 2a (no bias corr.)        & 1.308 & 0.869 & {2.949\,[.087]} & 0.990 & {1.093\,[.290]} & 0.976 \\
\quad Run 2b (weighted avg)         & 1.351 & 0.898 & {2.918\,[.113]} & 0.980 & {1.120\,[.499]} & 1.000 \\
\midrule
\multicolumn{7}{@{}l}{\textit{Panel D: Run~3 family}} \\[3pt]
Run 3 (final)                       & 1.216 & 0.808 & {3.244\,[.852]} & 1.089 & {1.305\,[.755]} & 1.165 \\
\quad Run 3a ($h{=}1$, no blend)    & 1.309 & 0.870 & {2.994\,[.648]} & 1.005 & {1.238\,[.818]} & 1.106 \\
\quad Run 3b ($h{=}2$, no blend)    & 1.307 & 0.869 & {3.037\,[.865]} & 1.020 & {1.276\,[.855]} & 1.140 \\
\bottomrule
\end{tabular}
\begin{tablenotes}\footnotesize
\item \textit{Notes:} RMSE of forecast errors for euro area real GDP growth (year on year). \rev{Search sample: 1999Q3--2016Q4 data; RMSE scored over 65 forecast dates (2000Q4--2016Q4).} Holdout: 2017Q1--2025Q4 (36~quarters). Holdout excl.\ COVID: holdout dropping 2020Q1--Q4 (32~quarters). Relative RMSE is computed against the simple average on each respective sample. Bracketed values in the holdout columns are one sided $p$-values from the Diebold--Mariano test that the method has superior predictive accuracy relative to the simple average, using the EWC fixed-$b$ approximation \citep{ShinSchor2026}; these have intrinsically low power given the small holdout sample (36~quarters, 32 excluding COVID) and serial correlation in forecast errors, and should be read as descriptive. peLASSO ex post (original) and best individual use parameters selected on the search sample and are evaluated without re-optimization on the holdout. peLASSO ex post (per window) re-optimizes $\lambda$ on each evaluation window separately.
\end{tablenotes}
\end{threeparttable}
\end{table}

\subsection{Role of individual components}

The intermediate methods isolate what drives generalization.

\paragraph{Stability selection (Run~1 vs.\ 1a, 1b).} The full Run~1 (holdout relative RMSE 0.945) substantially outperforms both alternatives: Run~1a (1.004), which uses elastic net selection without stability, and Run~1b (1.027), which uses standard peLASSO with CV. The stability selection mechanism and the composite performance weighting are jointly responsible for the holdout gain. Without them, elastic net PE selection is essentially no better than the simple average.

\paragraph{Bias correction (Run~2 vs.\ 2a, 2b).} Removing the bias correction from Run~2 degrades performance from 0.811 to 0.990 (Run~2a) or 0.980 (Run~2b). The bias correction contributes most of Run~2's advantage. However, Run~2a and 2b are still competitive with the simple average, indicating that the temporally weighted ranking and adaptive window mechanism also add modest value. The bias correction adjusts for persistent short run bias in combined forecast errors, which appears to reflect a structural feature rather than overfitting to the search sample.

\paragraph{Egalitarian elastic net blend (Run~3 vs.\ 3a, 3b).} Run~3's 70/30 blend with the egalitarian elastic net (holdout relative RMSE 1.089) performs worse than the simpler PE adaptive LASSO without the blend (3a at 1.005, 3b at 1.020). The ensemble component that improved search-sample fit introduced additional estimation noise that does not pay off on the holdout.

\subsection{Independent dataset robustness}

To verify that these results are not an artifact of using the original \texttt{H1\_gdp.csv} verbatim for the search-sample period, we also constructed a fully independent dataset where all 106~rows are built from downloaded SPF and Eurostat data, with no verbatim copy from the original file. The independent dataset uses the current (2026) GDP vintage rather than the 2018Q1 vintage in the original study; over the original sample the mean absolute revision is 0.079~pp and the maximum is 0.20~pp.

Table~\ref{tab:holdout_indep} repeats the exercise on this independent dataset. The key rankings are preserved: Run~2 achieves a relative RMSE of 0.808 on the holdout, Run~1 achieves 0.946, and the original paper methods achieve 0.974 and 0.968. Run~3's relative RMSE shifts from 1.089 to 0.962 on the independent dataset, but its $p$-values remain very large in both cases, so the difference is not meaningful. The independent dataset confirms that the main ranking is preserved across vintages: Run~2 strongest, Run~1 second, and Run~3 not distinguishable from the simple average.

\begin{table}[htbp]
\centering
\begin{threeparttable}
\caption{Full holdout evaluation: independent dataset}
\label{tab:holdout_indep}
\footnotesize
\begin{tabular}{@{}l S[table-format=1.3] S[table-format=1.3] S[table-format=1.3] S[table-format=1.3] S[table-format=1.3] S[table-format=1.3]@{}}
\toprule
 & \multicolumn{2}{c}{Search sample} & \multicolumn{2}{c}{Holdout} & \multicolumn{2}{c}{Holdout excl.\ COVID} \\
 \cmidrule(lr){2-3} \cmidrule(lr){4-5} \cmidrule(lr){6-7}
 & {RMSE} & {Relative} & {RMSE} & {Relative} & {RMSE} & {Relative} \\
\midrule
\multicolumn{7}{@{}l}{\textit{Panel A: Benchmarks and original paper methods}} \\[3pt]
Simple average          & 1.510 & 1.000 & 2.979 & 1.000 & 1.120 & 1.000 \\
peLASSO ex post (original)          & 1.417 & 0.938 & {3.041\,[.763]} & 1.021 & {1.350\,[.881]} & 1.206 \\
peLASSO ex post (per window)   & 1.417 & 0.938 & {2.976\,[.249]} & 0.999 & {1.119\,[.188]} & 0.999 \\
Best $\leq 6$-avg           & 1.500 & 0.994 & {2.901\,[.239]} & 0.974 & {1.043\,[.346]} & 0.931 \\
Best $(\leq 6, \leq 40)$-avg        & 1.298 & 0.860 & {2.884\,[.100]} & 0.968 & {1.094\,[.340]} & 0.977 \\
Best individual         & 1.407 & 0.932 & {2.909\,[.245]} & 0.977 & {1.109\,[.463]} & 0.990 \\
\midrule
\multicolumn{7}{@{}l}{\textit{Panel B: Run~1 family}} \\[3pt]
Run 1 (final)                       & 1.295 & 0.858 & {2.817\,[.177]} & 0.946 & {0.965\,[.233]} & 0.861 \\
\quad Run 1a (ENet, median)         & 1.479 & 0.980 & {2.991\,[.730]} & 1.004 & {1.157\,[.779]} & 1.033 \\
\quad Run 1b (peLASSO, CV)          & 1.455 & 0.964 & {3.051\,[.952]} & 1.024 & {1.250\,[.931]} & 1.116 \\
\midrule
\multicolumn{7}{@{}l}{\textit{Panel C: Run~2 family}} \\[3pt]
Run 2 (final)                       & 0.834 & 0.552 & {2.408\,[.127]} & 0.808 & {0.826\,[.107]} & 0.737 \\
\quad Run 2a (no bias corr.)        & 1.304 & 0.864 & {2.947\,[.090]} & 0.989 & {1.096\,[.298]} & 0.978 \\
\quad Run 2b (weighted avg)         & 1.343 & 0.890 & {2.898\,[.100]} & 0.973 & {1.065\,[.054]} & 0.950 \\
\midrule
\multicolumn{7}{@{}l}{\textit{Panel D: Run~3 family}} \\[3pt]
Run 3 (final)                       & 1.305 & 0.864 & {2.865\,[.309]} & 0.962 & {1.349\,[.728]} & 1.205 \\
\quad Run 3a ($h{=}1$, no blend)    & 1.490 & 0.987 & {2.960\,[.400]} & 0.994 & {1.270\,[.802]} & 1.133 \\
\quad Run 3b ($h{=}2$, no blend)    & 1.502 & 0.995 & {2.963\,[.414]} & 0.995 & {1.276\,[.812]} & 1.139 \\
\bottomrule
\end{tabular}
\begin{tablenotes}\footnotesize
\item \textit{Notes:} Same as Table~\ref{tab:holdout_full} but using the independently constructed dataset (all rows from downloaded SPF and Eurostat data, no verbatim copy from the original \texttt{H1\_gdp.csv}). Search-sample RMSEs differ from Table~\ref{tab:holdout_full} due to GDP vintage differences and independent imputation. Bracketed values in the holdout columns are one sided $p$-values from the Diebold--Mariano test using the EWC fixed-$b$ approximation \citep{ShinSchor2026}; see Table~\ref{tab:holdout_full} notes for details and caveats. peLASSO ex post (original) and best individual use parameters selected on the search sample; see Table~\ref{tab:holdout_full} notes for details.
\end{tablenotes}
\end{threeparttable}
\end{table}

\section{Illustration: beyond forecasting combination}
\label{sec:app_examples}

\rev{As discussed in Section~\ref{sec:fourfiles}, the} same four-file architecture applies whenever the researcher can separate an immutable evaluation design from an editable implementation rule. The remainder of this section provides three stylized illustrations that map the protocol into familiar empirical settings: regression specification search, structural VARs, and inflation forecasting for monetary policy.

These illustrations are intentionally schematic. In forecasting problems, the evaluator can often be a straightforward out-of-sample loss. In inference-heavy settings, by contrast, any scalar score should be understood as a diagnostic screen within a broader research design, not as a complete validation of identification or inference.

\paragraph{Example 1: Specification search in a regression model.}

Consider a researcher studying a descriptive wage regression using cross-sectional data. The estimating equation is
\[
\log w_i \;=\; \beta \cdot \text{educ}_i + g(z_i) + e_i,
\]
where $w_i$ is earnings, $\text{educ}_i$ is years of schooling, $\beta$ is a coefficient of interest, and $z_i$ is a vector of controls including experience, experience squared, demographic indicators, occupation, industry, and geographic region. The researcher wants to compare alternative specifications for the nuisance function $g(\cdot)$ while handling inference on $\beta$ separately.

\medskip\noindent\textit{File mapping.}

\begin{itemize}
\item \texttt{program.md} ($C$): ``Compare candidate specifications of $g(z_i)$. Only edit \texttt{train.R}. The search sample and its internal evaluation split are fixed in \texttt{prepare.R}. Do not modify the sample or the outcome variable.''

\item \texttt{train.R} ($\tau$): Defines a function that receives the training sample and returns fitted values for the nuisance component, potentially along with fitted values for $\text{educ}_i$ if a partialling-out design is used. The agent may explore polynomial terms, splines, interactions, LASSO-penalized specifications, or other functional forms for $g$.

\item \texttt{prepare.R} ($S$): Splits the search sample $D^S$ into a training partition $\mathcal{I}_{\text{train}}$ and an evaluation partition $\mathcal{I}_{\text{eval}}$, fixed ex ante. On $\mathcal{I}_{\text{train}}$, it sources \texttt{train.R} to obtain the nuisance fit, and if desired partials out both outcome and treatment before estimating $\hat{\beta}$. On $\mathcal{I}_{\text{eval}}$, it computes a holdout prediction error such as
\[
S(\tau; D^S) = \frac{1}{|\mathcal{I}_{\text{eval}}|} \sum_{i \in \mathcal{I}_{\text{eval}}} \bigl(\log w_i - \hat{\beta} \cdot \text{educ}_i - \hat{g}(z_i)\bigr)^2.
\]

\item \texttt{results.tsv} ($L_K$): Records each specification attempted, its evaluation MSE, and the implied $\hat{\beta}$.
\end{itemize}

If the scoring rule were the in-sample $p$-value for $\hat{\beta}$, the loop would amount to automated $p$-hacking. The discipline comes from fixing the evaluator and the sample split before the search begins and from disclosing the full search path. Causal claims about returns to schooling would still require separate identification assumptions. After the search, the researcher can evaluate the winning specification on a post-search holdout $D^H$ (e.g., a later survey year or a pre-reserved subsample withheld from the search) to check whether the nuisance fit generalizes.

\paragraph{Example 2: Structural VAR identification.}

Consider a three-variable VAR for macroeconomic analysis of monetary policy. Let $y_t = (\Delta \text{GDP}_t,\; \pi_t,\; i_t)'$, where $\Delta \text{GDP}_t$ is real GDP growth, $\pi_t$ is inflation, and $i_t$ is the federal funds rate. The reduced-form VAR is
\[
y_t = \Phi_1 y_{t-1} + \cdots + \Phi_p y_{t-p} + u_t,
\]
where $u_t$ is the reduced-form innovation with $\text{Var}(u_t) = \Sigma_u$. The structural representation is $u_t = A_0^{-1} \varepsilon_t$, where $\varepsilon_t$ is the vector of structural shocks and $A_0$ encodes the contemporaneous causal structure. The goal here is more modest: compare candidate ways of constructing a monetary policy shock series and record how each one lines up with an external narrative diagnostic.

\medskip\noindent\textit{File mapping.}

\begin{itemize}
\item \texttt{program.md} ($C$): ``Compare candidate constructions of a monetary policy shock series using a fixed external narrative diagnostic. Only edit \texttt{train.R}. Do not modify the VAR specification.''

\item \texttt{train.R} ($\tau$): Defines a function that receives the estimated reduced-form parameters $(\hat{\Phi}_1, \ldots, \hat{\Phi}_p, \hat{\Sigma}_u)$, together with any auxiliary diagnostic series supplied by \texttt{prepare.R}, and returns a candidate monetary policy shock series $\hat{\varepsilon}_t^{mp}$ and the associated GDP impulse response at horizon $h$. The agent may explore recursive orderings, sign restrictions, or other mappings from reduced-form innovations into candidate shock measures.

\item \texttt{prepare.R} ($S$): Estimates the reduced-form VAR on a fixed training sample, calls \texttt{train.R} to obtain the candidate shock, and computes a diagnostic score on $D^S$:
\[
S(\tau; D^S) = -\left|\text{corr}\!\left(\hat{\varepsilon}_t^{mp},\; z_t^{RR}\right)\right|,
\]
where $z_t^{RR}$ is the Romer and Romer narrative monetary policy shock measure. The negative sign ensures that higher correlation corresponds to a lower (better) score. This score is best read as an external diagnostic rather than a standalone proof of identification.

\item \texttt{results.tsv} ($L_K$): Records the candidate construction, the diagnostic score, and the implied peak response of GDP to a monetary policy shock.
\end{itemize}

The point is not that the agent ``finds the largest effect'' or settles identification, but that any search over candidate shock constructions is visible, bounded, and judged against a criterion fixed ex ante. After the search, the researcher can evaluate the chosen shock construction on a later sample period $D^H$ (e.g., search on 1969--2000, holdout on 2001--2007) to check whether the narrative correlation and impulse response patterns hold outside the search sample.

\paragraph{Example 3: Inflation forecasting for monetary policy.}

Central banks routinely produce inflation forecasts to guide monetary policy decisions. The workhorse framework is the Phillips curve:
\[
\pi_{t+h} = \alpha + \beta \cdot \text{slack}_t + \gamma \cdot \pi_t^e + \delta' z_t + e_{t+h},
\]
where $\pi_{t+h}$ is $h$-quarter-ahead CPI inflation, $\text{slack}_t$ is a measure of economic slack, $\pi_t^e$ is an inflation expectations proxy, and $z_t$ is a vector of additional conditioning variables (commodity prices, exchange rates, financial conditions). The practical difficulty is that the specification involves many choices: which slack measure (unemployment gap, output gap, capacity utilization), which expectations proxy (survey expectations, breakeven inflation, adaptive expectations), which additional predictors, which estimation method (OLS, ridge, random forest), and which estimation window.

\medskip\noindent\textit{File mapping.}

\begin{itemize}
\item \texttt{program.md} ($C$): ``Compare candidate real-time Phillips curve specifications for $h = 4$ quarter-ahead CPI inflation forecasting. No lookahead: at forecast origin $t$, only real-time vintage data through $t$ is available. Only edit \texttt{train.R}. Budget: $K = 200$ experiments.''

\item \texttt{train.R} ($\tau$): Defines a function that receives the information set $\mathcal{F}_t = \{y_s, x_s\}_{s=1}^{t}$ and returns a point forecast $\hat{\pi}_{t+h|t}$. The agent may explore different slack measures, expectations proxies, transformations, variable selection methods, penalized estimation, or nonlinear specifications.

\item \texttt{prepare.R} ($S$): Implements pseudo-out-of-sample evaluation. For each forecast origin $t \in \{t_0, t_0 + 1, \ldots, T-h\}$, passes the real-time information set to \texttt{train.R} and collects the forecast. The score is
\[
S(\tau; D^S) = \text{RMSE} = \left(\frac{1}{|\mathcal{E}|}\sum_{t \in \mathcal{E}} \bigl(\pi_{t+h} - \hat{\pi}_{t+h|t}\bigr)^2\right)^{1/2}.
\]

\item \texttt{results.tsv} ($L_K$): Records each Phillips curve specification, its RMSE, and a description of the method (e.g., ``ridge Phillips curve, unemployment gap, SPF expectations, 40-quarter rolling window'').
\end{itemize}

This is a natural fit for the protocol because the evaluator is a genuine pseudo-out-of-sample forecast loss. For central bank staff who routinely compare Phillips curve specifications, the protocol would make the search path more visible and easier to disclose. The researcher can further extend the evaluation to a later period $D^H$ not available during the search to check whether the selected specification generalizes.

\bigskip
These three illustrations suggest how the four-file protocol can be adapted across settings. The architecture applies whenever the researcher can separate an immutable evaluation design from an editable implementation rule. Forecasting is the cleanest instance because the score is naturally out-of-sample. In more inference-heavy settings, the protocol is better viewed as a way to make adaptive search more visible and easier to disclose, not as a substitute for the underlying econometric argument.

A post-search holdout can also help reduce concerns about p-hacking, but only in a limited sense. Its value is that it separates discovery from evaluation: the agent may search adaptively over many specifications on the search sample, but the selected output is then judged on data not used during that search. This makes sample-specific improvements less likely to be mistaken for robust findings. At the same time, a holdout is not a complete solution. If the holdout is consulted repeatedly and the method is revised in response, it becomes part of the search process itself. And in inference-heavy settings, holdout performance on a diagnostic criterion does not by itself validate identification, causal interpretation, or post-selection inference. The holdout should therefore be understood as a practical guardrail against hidden adaptive search, not as a full cure.

\section{Formal statement of the search problem}
\label{sec:app_formal}

\rev{This appendix restates the protocol of Section~\ref{sec:protocol} as a search problem and records, in standard terms, the distinctions that the main text states in plain language.}

\rev{Let $C$ denote the written instruction contract and let $\mathcal{T}(C)$ denote the class of editable scripts admissible under $C$. Let the available data be partitioned as $D = (D^{S}, D^{H})$, where $D^{S}$ is the search sample and $D^{H}$ is a post-search holdout reserved for external validation. Let $S(\tau; D^{S})$ denote the scalar score returned by the immutable evaluator when script $\tau \in \mathcal{T}(C)$ is run on the search sample. Both $\mathcal{T}(C)$ and $S$ are fixed ex ante by the researcher as part of the search design. When lower scores are preferred, the idealized search problem is
\[
\tau^{\star} \in \arg\min_{\tau \in \mathcal{T}(C)} S(\tau; D^{S}).
\]
In practice, the agent sequentially generates and evaluates a finite sequence of candidate scripts $\{\tau_{0}, \tau_{1}, \ldots, \tau_K\}$, each informed by the outcomes of previous attempts, and returns
\[
\hat{\tau}_{K} \in \arg \min_{\tau \in \{\tau_{0}, \tau_{1}, \ldots, \tau_{K} \}} S(\tau; D^{S}).
\]
The distinction matters: $\tau^{\star}$ is the best admissible script in the full class $\mathcal{T}(C)$, while $\hat{\tau}_K$ is the best script found by one bounded, path dependent search. The audit log records the outcome of every experiment as an ordered sequence $L_K = \bigl((k, \tau_k, s_k, d_k)\bigr)_{k=0}^{K}$, where $s_k = S(\tau_k; D^{S})$ when the run succeeds, $s_k = +\infty$ when it crashes, and $d_k \in \{\textit{keep},\, \textit{discard},\, \textit{crash}\}$ is the outcome status defined in Section~\ref{sec:fourfiles}. Once the search is complete, the researcher may evaluate the selected script on the holdout, $S^{H}(\hat{\tau}_K; D^{H})$, where $S^{H}$ is a scoring function computed on $D^{H}$; nothing in the loop requires $D^{H}$ to exist during the search.}

\rev{Two distinctions used in the main text can now be made precise. First, a restriction in the contract is \emph{machine checkable} if a violation can be detected mechanically, without judgment. The evaluator enforces the information-set restriction by controlling which data are passed to the script. Other restrictions, including the editable file, runtime limit, and experiment budget, are mechanically auditable from the run logs, git history, and run counts, rather than actively prevented in this implementation. A restriction such as ``do not combine peLASSO with other methods'' is not machine checkable: any syntactically valid forecasting script runs and receives a score, whatever method it implements. Second, the \emph{effective admissible set} $\mathcal{T}(C)$ is the set of scripts that satisfy the machine checkable restrictions. It may be strictly larger than the set of scripts the researcher had in mind when writing the contract, which we call the researcher's \emph{semantic intent}. In conventional specification search the admissible set is an explicit menu written down in advance, so the effective set and the semantic intent coincide by construction; in agentic search over programs they can diverge. Drift, as used in Section~\ref{sec:results}, is a search path that remains inside the effective admissible set while leaving the semantic intent. The protocol makes containment in the first set verifiable and discloses the entire path; it does not, and cannot, guarantee containment in the second. The post-search holdout is the instrument that evaluates whether a departure from the semantic intent generalizes beyond the search sample.}

\end{document}